\documentclass[twocolumn,showpacs,preprintnumbers,amsmath,amssymb]{revtex4}%
\usepackage{graphicx}
\usepackage{dcolumn}
\usepackage{bm}
\usepackage{amsmath}
\usepackage{amsfonts}
\usepackage{amssymb}%
\begin{document}
%\preprint{APS/123-QED}
\title{Error correction in ensemble registers for quantum repeaters and quantum computers}
\author{Etienne Brion$^{1,2}$, Line Hjortsh\o j Pedersen$^3$, Mark Saffman$^4$, and Klaus M\o lmer$^3$}
\affiliation{$^{1}$ Institute for Mathematical Sciences, Imperial
College London, SW7 2PE, UK} \affiliation{$^{2}$ QOLS, Blackett
Laboratory, Imperial College London, SW7 2BW, UK}
\affiliation{$^{3}$Lundbeck Foundation Theoretical Center for
Quantum System Research, Department of Physics and Astronomy,
University of Aarhus, DK-8000 \AA rhus C, Denmark}
\affiliation{$^{4}$Department of Physics, University of Wisconsin,
1150 University Avenue, Madison, Wisconsin 53706, USA}
\date{\today}

\begin{abstract}
We propose to use a collective excitation blockade mechanism to identify errors that occur due to
disturbances of single atoms in ensemble quantum registers where qubits are stored in the collective
population of different internal atomic states. A simple error correction procedure and a simple
decoherence-free encoding of ensemble qubits in the hyperfine states of alkali atoms are presented.

\end{abstract}
\maketitle

The potential of quantum computing, e.g., for factoring and
unstructured search \cite{nielsenchuang}, lies in the significant
speed-up in comparison with classical computing allowed by the use
of superposition states. Scalability of quantum devices to
operations on a significant number of bits is crucial for all
quantum computer proposals and is in principle achieved when
information is stored in binary form in physically distinct
two-level quantum systems since any unitary transformation on the
full register product space can then be generated by a suitable set
of single- and two-qubit operations. In atomic quantum computing
proposals, for instance \cite{ions,jaksch,saffman}, bits are encoded
in different atoms or ions, the collection of which forms a quantum
register. In this case, scalability is not limited by any shortage
of atoms in laboratory experiments, but by the immense difficulty of
preserving and manipulating the quantum state of such a
multi-component
system.\\
\indent As an alternative to storage and encoding of quantum
information in single quantum systems, it has been proposed to use
atomic ensembles, which have favorably enhanced interactions with
optical fields \cite{dlcz} and long coherence times despite the fact
that even a single qubit is stored in a very entangled state of the
many-atom system \cite{mewes}. Encoding and storage of qubits in
ensembles is a promising approach to repeater technologies for long
distance quantum communication \cite{dlcz,kimble}, and recently, we
proposed an approach towards scalability of quantum computing which
combines the ensemble ideas and the rich internal level structure
found in many quantum systems \cite{brion}. Rather than storing
individual bits in individual particles, we suggest to encode
quantum information in the symmetric Fock space describing the
collective population of the different internal states of the
particles in the ensemble.
%To be more specific, we proposed to physically
%encode an arbitrary $N$-bit register state $\left\vert b_{1}b_{2}...b_{N}\right\rangle $, $b_{i}=0,1$,
%in the collective symmetric state of $K$ identical quantum systems with $\left( N+1\right)  $ internal
%levels $\left\{ \left\vert i\right\rangle ,i=0,\ldots,N\right\}  $, in which $b_{i}$ atoms populate
%the internal state $\left\vert i\right\rangle $, while $K-\sum_{i=1}^{N}b_{i}$\ particles are in the
%so-called reservoir state $\left\vert 0\right\rangle $.
One- and two-bit quantum gates as needed for quantum computing and for entanglement swapping and
purification of a quantum repeater station based on the atomic multi-level structure may now be
carried out by operations
within the space of symmetric collective states of a single atomic ensemble.\\
\indent An experimental difficulty consists in restricting the occupation of the internal states to
the bit values zero and unity as resonant driving of independent particles on an internal state
transition naturally leads to multiple occupation of the same internal state. In atomic systems,
however, one can force the system to remain in this desired subspace by use of the Rydberg blockade
mechanism \cite{jaksch,lukin}. This phenomenon is due to the strong long-range interaction between
highly excited atoms and the associated energy shifts, which prevent atoms in the vicinity of an
excited neighbor from being transferred into a Rydberg state. Taking advantage of this mechanism one
can drive precisely one atom from the initially macroscopically populated state $|s\rangle$, via a
Rydberg state $|r\rangle$, into an initially empty internal state $|i\rangle$. One can also perform
arbitrary single-qubit rotations \cite{lukin}, by \textit{(a)} mapping $|i\rangle$ to $|r\rangle$ by a
$\pi$-pulse, \textit{(b)} coupling the states with none and a single Rydberg excited atom for
adjustable amounts of time, before \textit{(c)} mapping the $|r\rangle$-amplitude back onto
$|i\rangle$ with a final $\pi$-pulse. Since the excitation to a Rydberg level from one of the internal
states can prevent excitation from another internal state, two-qubit gates can be implemented through
the same mechanism \cite{brion}. It is important to point out, that the operations mentioned are
carried out by collective addressing, and the exciting laser pulses only need to be adjusted to
the energies and coupling strengths of the atomic states. \\
\indent So far, we have described how the ensemble encoding works
for qubit storage or computing under ideal conditions. For our
proposal to be practically useful, we must address the errors which
may occur and devise methods to repair them. Error correction
techniques, derived under the assumption that errors occur
independently on separate qubits, do not apply here since we work
under the restriction that we can only collectively and
symmetrically address the ensemble. We shall show, however, that
even under these circumstances efficient error identification and
correction is possible in our framework.
\begin{figure}
\centering
\includegraphics[width=4cm]{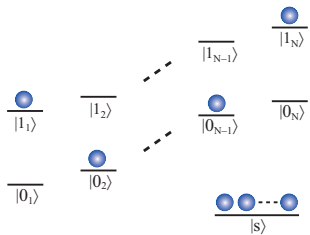}
\caption{(color online) Qubit encoding in the symmetric states of an ensemble of
$(2N+1)$-level systems. $|s\rangle$ denotes the reservoir state. The
figure depicts the state $|10\ldots 01\rangle$.}
\label{levels}
\end{figure}
\\ \indent Let us first present an encoding scheme, which differs from the one proposed in \cite{brion},
but which will greatly simplify our error analysis. We consider an
ensemble of $K$ atoms with $2N+1$ internal states, denoted
$\left\vert s\right\rangle ,\left\vert 0_{i}\right\rangle
,\left\vert 1_{i}\right\rangle ,$ $i=1,...,N$ (see Fig.
\ref{levels}). We assume that the system can be initially prepared
in the reservoir product state $|S\rangle_{K}\equiv|ss..s\rangle$.
We also suppose that the different atomic transitions $\left\vert
0_{i}\right\rangle \longleftrightarrow\left\vert 1_{i}\right\rangle
$, $i=1,\ldots,N$, can be driven selectively. Using the Rydberg
blockade we produce the symmetric state of the ensemble with
precisely one atom transferred from $\left\vert s\right\rangle $
into each of the $\left\vert 0_{i}\right\rangle $ states, which
encodes the logical state $|0_{1},0_{2},\ldots,0_{N}\rangle$. In the
following, we shall only apply Hamiltonians which leave the system
in the symmetric Fock subspace corresponding to a single occupation
of each manifold $\left\{ \left\vert 0_{i}\right\rangle ,\left\vert
1_{i}\right\rangle \right\}  $, $i=1,\ldots,N$, at least until
errors occur. One-bit rotations on qubit $i$ are achieved by driving
the single-atom transition $\left\vert 0_{i}\right\rangle
\leftrightarrow\left\vert 1_{i}\right\rangle $, without involving
excitation to Rydberg states and are thus faster and easier to
implement than in \cite{lukin,brion}. Conditional two-bit dynamics,
however, still relies on long-range Rydberg interactions by using a
$\pi$-pulse, e.g., $\left\vert 1_{i}\right\rangle
\rightarrow\left\vert r\right\rangle $, to conditionally prevent
another qubit $j\neq i$, from being excited
into $\left\vert r\right\rangle $.\\
\indent Due to interaction with their local environment (background gas collisions, spontaneous
emission of radiation), it is physically motivated to assume that the individual particles which form
a quantum register are affected by independent errors. In standard encoding schemes, where qubits are
stored in separate particles, this assumption implies that qubits are corrupted independently of each
other. Error correction schemes, based on encoding logical bits in several physical qubits, then
consist in syndrome measurement and conditional back action which corrects such errors if they are not
too frequent \cite{steane}. In our ensemble encoding approach, the physical error model implies that
the system may leave the symmetric, computational subspace. In the following paragraphs, we review the
different types of errors which can affect
the system and show how to correct them.\\
\indent The first error source we shall consider is atom loss. For
the sake of simplicity, we start with the case of a one-bit
register, prepared in the logical 0 state, denoted
$|\overline{0}\rangle_{K}$, to recall the symmetric collective
character and the number $K$ of atoms contributing to the ensemble
state,
\begin{align}
\left\vert \overline{0}\right\rangle _{K} &  =\frac{1}{\sqrt{K}}\left(
\left\vert 0ss\ldots s\right\rangle +\left\vert s0s\ldots s\right\rangle
+\ldots+\left\vert ss\ldots s0\right\rangle \right)  \nonumber\\
&  =\frac{1}{\sqrt{K}}\left\vert 0\right\rangle \otimes\left\vert
S\right\rangle _{K-1}+\sqrt{\frac{K-1}{K}}\left\vert s\right\rangle
\otimes\left\vert \overline{0}\right\rangle _{K-1}.\label{onebit}%
\end{align}
The removal of any, say the first, atom, produces the state
\begin{equation}
\left\vert \psi\right\rangle =\frac{1}{\sqrt{K}}|S\rangle_{K-1}+\sqrt
{\frac{K-1}{K}}|\overline{0}\rangle_{K-1},\label{rem}%
\end{equation}
or a mixed state with equivalent weight factors on the two
components. For large $K$, this state is dominated by the second
term, which encodes the correct register state but with a smaller
total number of atoms in the ensemble. Since gates act on this state
in the same way as on the state encoded with the original number of
atoms, the error due to the atomic loss is therefore only connected
with the first component in (\ref{rem}), which will be propagated by
the subsequent unitary dynamics and introduce a very small
probability (i.e., the probability that the loss occurred multiplied
with the factor $1/K$) for an erroneous output at the end of the
calculation. The same reasoning can be applied to any qubit
superposition state yielding the same result, and a simple
calculation shows that in an $N$ bit register, loss of a single atom
introduces an erroneous component with population $\frac{N}{K}$.
%the extension to an $N$-bit register follows
%from the generalization of Eq.(\ref{onebit})
%to the $N$-bit ensemble state $\left\vert
%\overline{b}_{1}\overline{b}_{2}\ldots\overline{b}_{N}\right\rangle _{K}$, $\left(  b_{i}=0,1\right)$:
%the loss of, say, the first atom yields the state
%\begin{align}
%\left\vert \psi\right\rangle  &  =\frac{1}{\sqrt{K}}\left(  \left\vert
%\overline{b}_{2}\ldots\overline{b}_{N}\right\rangle _{K-1}+\left\vert
%\overline{b}_{1}\overline{b}_{3}\ldots\overline{b}_{N}\right\rangle
%_{K-1}\right.  \nonumber\\
%&  \left.  +\ldots+\left\vert \overline{b}_{1}\ldots\overline{b}%
%_{N-1}\right\rangle _{K-1}\right)  \nonumber\\
%&  +\sqrt{\frac{K-N}{K}}\left\vert \overline{b}_{1}\ldots\overline{b}%
%_{N}\right\rangle _{K-1}.\label{Nerrorstate}%
%\end{align}
%thus
%introducing an error with the probability
For very large $K$ we may accept the erroneous component in the wave function, or we may identify the
error, by carrying out a measurement to find out if one of the subspaces $\{|0_{i}\rangle,\
|1_{i}\rangle\}$ is not populated. By transferring an atom from $|s\rangle$ to that subspace via the
Rydberg state, we reestablish a legal, but most likely erroneous, register state. The error can,
however, be addressed by suitable error correction techniques, which
we shall outline below.\\
%%
%\indent Note that the feeding of the register states from the reservoir level $|s\rangle$ involves a
%$\pi$ pulse transfer to the collectively excited Rydberg state, with a Rabi frequency which is
%enhanced by the large number of atoms occupying the reservoir state. Imprecise knowledge of the
%precise number of  atoms available for the transition may be compensated by use of composite pulse
%techniques \cite{composite}.\\
%
\indent Atom loss is not the most critical error to affect the ensemble since the resulting state
retains the full permutation symmetry, and the computation may go on safely. It is more problematic
when an atom is disturbed and remains in the sample where it continues to interact with the other
atoms and corrupts the future quantum gates.\\
\indent For the sake of simplicity, let us examine the case of a
single-bit register initially prepared in the single-qubit state,%
\begin{align}
\alpha\left\vert \overline{0}\right\rangle _{K}+\beta\left\vert \overline
{1}\right\rangle _{K} &  =\frac{1}{\sqrt{K}}\left(  \alpha\left\vert
0\right\rangle +\beta\left\vert 1\right\rangle \right)  \otimes\left\vert
S\right\rangle _{K-1}\nonumber\\
&  +\sqrt{\frac{K-1}{K}}\left\vert s\right\rangle \otimes\left(
\alpha\left\vert \overline{0}\right\rangle _{K-1}+\beta\left\vert \overline
{1}\right\rangle _{K-1}\right)  ,\label{state}%
\end{align}
If the first atom is affected by an error, the term in the first line has negligible amplitude
compared to the term in the second line if $K$ is large, and it will be neglected in the following.
Suppose that the $|s\rangle$ state of the first atom evolves into
\begin{equation}
\left\vert \phi\right\rangle =c_{0}\left\vert 0\right\rangle +c_{1}\left\vert
1\right\rangle +c_{s}\left\vert s\right\rangle +\left\vert \phi^{\prime
}\right\rangle \label{diststate}%
\end{equation}
where $\left\vert \phi^{\prime}\right\rangle $ is orthogonal to $\left\vert 0\right\rangle ,\left\vert
1\right\rangle ,$ and  $\left\vert s\right\rangle $, and that the resulting erroneous state of the
ensemble thus takes the form
\begin{align}
\left\vert \psi_{er}\right\rangle  &= \sqrt{\frac{K-1}{K}}\left(
c_{s}\left\vert s\right\rangle +\left\vert
\phi^{\prime}\right\rangle \right)  \otimes\left(  \alpha\left\vert
\overline{0}\right\rangle _{K-1}+\beta\left\vert
\overline{1}\right\rangle
_{K-1}\right)  \nonumber\\
&  +\sqrt{\frac{K-1}{K}}\left(  c_{0}\left\vert 0\right\rangle +c_{1}%
\left\vert 1\right\rangle \right)  \otimes\left(  \alpha\left\vert
\overline{0}\right\rangle _{K-1}+\beta\left\vert
\overline{1}\right\rangle _{K-1}\right)  .
\label{erstate}%
\end{align}
In the first line, the $\left\vert \phi^{\prime }\right\rangle $
component will never interfere with the computation, and the single
atom state $\left\vert s\right\rangle $ couples so weakly to our
laser driving fields in comparison with the symmetric states of the
remaining atoms that it will only slightly perturb the state of the
ensemble. This robustness against perturbations on individual atoms
is crucial for many uses of ensembles \cite{mewes}, and has been
crucial in experiments, e.g., on continuous variable quantum storage
in atomic ensembles, which already in their initial state may have a
significant number of atoms that are not pumped into the symmetric
state \cite{polzik}. The second line in Eq.(\ref{erstate}), however,
shows dangerous double occupancy of register states $0,1$ which must
be suppressed by physical manipulation of the system.
\begin{figure}
\centering
\includegraphics[width=5.5cm]{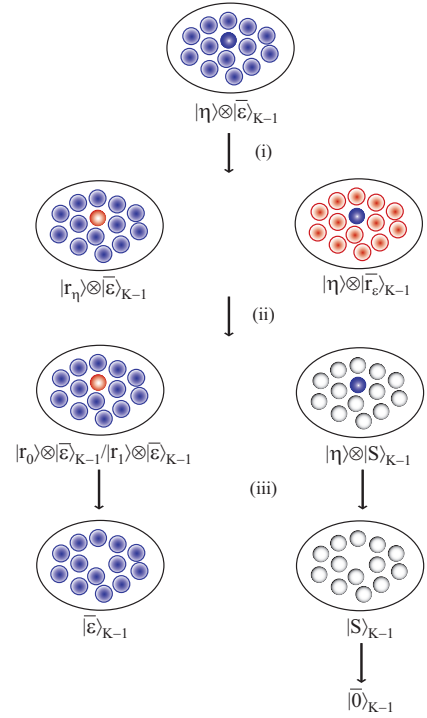}
\caption{(color online) Schematic illustration of our procedure to remove single
atom occupancy of register states, already occupied by the ensemble.
The second and the third line of the figure depict superposition
states. We use the short hand notation $|\eta\rangle = c_0|0\rangle
+ c_1|1\rangle$, $|r_\eta\rangle = c_0|r_0\rangle+c_1|r_1\rangle$,
$\left\vert \epsilon\right\rangle =
\alpha|\overline{0}\rangle_{K-1}+\beta|\overline{1}\rangle_{K-1}$
and $\left\vert r_\epsilon \right\rangle =
\alpha|\overline{r}_0\rangle_{K-1}+\beta|\overline{r}_1\rangle_{K-1}$.
Our error detection sequence proceeds as follows: $(i)$
$0\leftrightarrow r_0$ and $1\leftrightarrow r_1$ resonant composite
pulses are applied. $(ii)$ After detection of either $|r_0\rangle$
or $|r_1\rangle$, either an $s\leftrightarrow r_0$ or an
$s\leftrightarrow r_1$ resonant pulse is applied. $(iii)$ Depending
on the result of the measurement in $(ii)$ either $|r_0\rangle$ or
$|r_1\rangle$ is ionized. If an ion signal is observed, the final
state is $|\overline{\epsilon}\rangle_{K-1}$, otherwise
$|\eta\rangle$ is ionized yielding the state $|S\rangle_{K-1}$,
subsequently transferred into the final state
$|\overline{0}\rangle_{K-1}$ by an $s\leftrightarrow 0$ resonant
pulse.} \label{scheme}
\end{figure}
\\ \indent To this end, we apply the following procedure, which is
illustrated in Fig. \ref{scheme}. We first simultaneously apply two laser beams coupling $\left\vert
0\right\rangle $ and $\left\vert 1\right\rangle $\ to two different Rydberg states $\left\vert
r_{0}\right\rangle $\ and $\left\vert r_{1}\right\rangle $ with the same Rabi frequency $\Omega$. We
assume that both Rydberg states $\left \vert r_0 \right \rangle$ and $\left \vert r_1 \right \rangle$
block multiple Rydberg excitations. As a consequence, the states $\left\vert x\right\rangle
\otimes\left\vert \overline{y}\right\rangle _{K-1}$, with $x=s,\phi^{\prime}$ and $y=0,1$, couple to
the states $\left\vert x\right\rangle \otimes\left\vert \overline{r}_{y}\right\rangle _{K-1}$ with the
same coupling strength $\hbar\Omega$, while the states with $x=0,1$ and $y=0,1$, couple to $\left(
\left\vert r_{x}\right\rangle \otimes\left\vert \overline{y}\right\rangle _{K-1}+\left\vert
x\right\rangle \otimes\left\vert \overline{r}_{y}\right\rangle _{K-1}\right) /\sqrt{2}$, with the
coupling strength $\sqrt{2}\hbar\Omega$. Thanks to the second coupling strength being larger by a
factor of $\sqrt{2}$, it is possible to design a composite pulse sequence \cite{linespulse} which
leaves the first line in Eq.(\ref{erstate}) (and also any non-erroneous state) unchanged while
transforming the second line into
\begin{equation*}
\begin{split}
& \sqrt{\frac{K-1}{2K}} \Bigl[ (c_0 |r_0\rangle + c_1 |r_1\rangle)
\otimes\left( \alpha |\overline{0}\rangle_{K-1}+ \beta
|\overline{1}\rangle_{K-1}\right)  \\
& + \left( c_0 |0\rangle + c_1 |1\rangle \right)\otimes \left(
\alpha |\overline{r}_0\rangle_{K-1} + \beta
|\overline{r}_1\rangle_{K-1}\right) \Bigr]
\end{split}
\end{equation*}
We can now check whether an error has occurred or not by measuring
the Rydberg state content in the ensemble, e.g., by means of the
blockade of a neighboring read-out ensemble \cite{mark}. Such a
measurement is projective and if a Rydberg excitation in $\left\vert
r_{0}\right\rangle $ is detected we get the unnormalized state
\begin{align*}
&  c_{0}\left\vert r_{0}\right\rangle \otimes\left(  \alpha\left\vert
\overline{0}\right\rangle _{K-1}+\beta\left\vert \overline{1}\right\rangle
_{K-1}\right)  \\
&  +\alpha\left(  c_{0}\left\vert 0\right\rangle +c_{1}\left\vert
1\right\rangle \right)  \otimes\left\vert \overline{r}_{0}\right\rangle
_{K-1}.
\end{align*}
The goal is now to modify this state so that it can be used for
further processing. Since the coupling of the symmetrically excited
Rydberg state $\left\vert \overline{r}_{0}\right\rangle _{K-1}$ to
the reservoir state $|S\rangle_{K-1}$ is $\sqrt{K-1}$ times larger
than the coupling of the single atom states $|r_{0}\rangle$ and
$|s\rangle$, a resonant $\pi$-pulse on the $s\leftrightarrow r_{0}$
transition can be driven, which makes a complete transfer of the
collective state $|\overline{r}_0\rangle$ into $|S\rangle$ and only
a $\sim1/\sqrt{K}$ transfer of the single atom component
$|r_0\rangle$
into $|s\rangle$ \cite{linespulse2}.\\
\indent After such a pulse we therefore mainly obtain
\begin{align*}
&  c_{0}\left\vert r_{0}\right\rangle \otimes\left(  \alpha\left\vert
\overline{0}\right\rangle _{K-1}+\beta\left\vert \overline{1}\right\rangle
_{K-1}\right)  \\
&  +\alpha\left(  c_{0}\left\vert 0\right\rangle +c_{1}\left\vert
1\right\rangle \right)  \otimes\left\vert S\right\rangle _{K-1}.
\end{align*}
We then apply an ionizing pulse of the Rydberg state $\left\vert r_{0}\right\rangle $. If an ion is
observed we retain the correct initial qubit state $\left(  \alpha\left\vert \overline{0}\right\rangle
_{K-1}+\beta\left\vert \overline{1}\right\rangle _{K-1}\right)  $ with one atom less in the ensemble.
If no ion is detected, we obtain $\left( c_{0}\left\vert 0\right\rangle +c_{1}\left\vert
1\right\rangle \right) \otimes\left\vert S\right\rangle _{K-1}$. One then applies pulses ionizing the
states $\left\vert 0\right\rangle $ and $\left\vert 1\right\rangle $, which finally leads to
$\left\vert S\right\rangle _{K-1}$. We complete the procedure by applying a resonant pulse
$s\leftrightarrow0$ so that $\left\vert S\right\rangle _{K-1}$ is transferred into $\left\vert
\overline{0}\right\rangle _{K-1}$.\\
\indent If, during the first Rydberg state measurement, $\left\vert r_{0}\right\rangle $ is not
observed,  and if $\left\vert r_{1}\right\rangle $ is subsequently detected, one can apply similar
procedures as above. If neither $\left\vert r_{0}\right\rangle $ nor $\left\vert r_{1}\right\rangle $
are detected, the system is projected onto a (sufficiently) non-erroneous component in
Eq.(\ref{erstate}) that the ensemble still reliably
encodes the desired register state.\\
\indent By applying the detection/correction procedure described
above to a single-qubit register, we either get the correct initial
state $\left( \alpha\left\vert \overline{0}\right\rangle
_{K-1}+\beta\left\vert \overline {1}\right\rangle _{K-1}\right)  $
or end up in $\left\vert \overline{0}\right\rangle _{K-1}$. It
should be emphasized that as long as no error or a simple particle
loss occur, no error signal is reported by the Rydberg populations,
and no atoms are unnecessarily removed from the ensemble. In an
$N$-bit register one merely has to check all the qubit populations
successively as described above. At the end, we know whether and
where an error has occurred. If needed, we can repair the state of
the erroneous qubit, provided we encode each bit of information in a
superposition state $\alpha(|\overline{0}\overline {0}\rangle
_{K}+|\overline{1}\overline{1}\rangle _{K})/\sqrt{2}+\beta(|
\overline{0}\overline{1}\rangle _{K}+|
\overline{1}\overline{0}\rangle _{K})/\sqrt{2}$, such that two
ensemble qubits encode one logical qubit. Going through the above
error identification and register restoration scheme, assuming an
error has happened in any of the two register positions, we either
recover the same superposition, but with one atom less in the
ensemble, or superpositions involving only the first or the second
component of our two-bit code words, such as $\alpha|
\overline{0}\overline {0}\rangle _{K-1}+\beta|
\overline{0}\overline{1}\rangle _{K-1}$, from which the entire code
words can be reconstructed by simple gate operations. This
error-correction encoding is simpler than in the usual tensor
product encoding because our error identification protocol provides
direct information about which qubit has to be repaired. We
anticipate that other schemes may be derived to fix errors without
ionization of the atoms, but the reduction in size of the symmetric
ensemble due to errors seems unavoidable  We note, however, that it
is in principle possible to transfer the quantum state of our
diminished ensemble by the Rydberg blockade mechanism to a nearby
independent ensemble of
atoms, and such new ensembles can be supplied when needed.\\
\indent If the qubits are encoded in the internal Zeeman sublevels
of the lower atomic hyperfine levels in alkali atoms with ground
hyperfine manifolds $\{f,f+1\}$, a bias magnetic field along the
$z$-axis both makes the system immune to fluctuating orthogonal $x$-
and $y$-components of the field and provides an energy splitting of
the qubit states so that they can be unambiguously addressed by
resonant laser fields \cite{brion}. Fluctuations of the
$z$-component of the field perturb the atomic energy levels, but we
can make the system immune to these fluctuations if qubits are
associated with the pairs of states
$\{|f,m_{f}\rangle,|f+1,-m_{f}\rangle\}$, with
$m_{f}=0,\pm1,\pm2,..\pm f$. The states
$|f,m_{f}\rangle,|f+1,-m_{f}\rangle$ experience the same linear
Zeeman shift and only  small differential quadratic shifts in
the presence of magnetic field fluctuations, which therefore only
slightly disturb the relative phase of qubit states 0 and 1.
Coherence times exceeding one second have thus been observed for
superpositions of such states with $m_f=\pm 1$
in $^{87}$Rb \cite{treutlein}.\\
\indent In conclusion, we have proposed a novel scheme for encoding
multiple bits of quantum information in symmetric collective states
of an ensemble of multilevel atoms by making effective use of the
single particle Hilbert space dimension. Errors on individual atoms
can be repaired successfully if the ensemble is large enough. We
furthermore showed that a specific choice of qubit levels provides
an automatic decoherence-free subspace encoding against fluctuations
in the external magnetic field acting on all atoms. We presented the
analysis for the special case of a global Rydberg blockade
mechanism, but we anticipate that the scheme presented here for
identifying and eliminating systems which have left the symmetric
subspace may apply also with other effective interactions as a
general restoration mechanism for symmetric states in ensembles of
identical particles involved in quantum computing, long term quantum
memories and in quantum repeaters.

This work was supported by ARO-DTO Grant No. 47949PHQC and the European Union Integrated Project
SCALA.


\begin{thebibliography}{9}
\bibitem{linespulse} Characterizing a pulse by $(\theta,\phi)$,
where $\theta=\Omega t$ and $\phi$ are the pulse area and phase, the pulse sequence
$(\pi,0),(2\pi,\phi'),(\pi,0)$, where $\phi' = \arccos (\cot^2(\sqrt{2}\pi))$, implements $I$ for a
coupling strength of $\Omega$ and $\sigma_y$ (up to phase factors) for $\sqrt{2}\Omega$.
\bibitem{linespulse2} The erroneous amplitude transfer scaling as $\sim 1/K^{1/2}$ reduces to
$\sim 1/K^{3/2}$ if the $\pi$-pulse on the desired strong transition is replaced by three consecutive
$\pi$-pulses with phases $0,2\pi/3,4\pi/3$.
\end{thebibliography}
\end{document}